\documentclass[prb,notitlepage]{revtex4-1}

\usepackage{amsmath}   % need for subequations
\usepackage{graphicx}   % for figures
\usepackage{color}
\usepackage{verbatim}
\usepackage{amssymb}
\usepackage{epsfig}

\usepackage[dvipsnames]{xcolor}

\usepackage{cancel}
\usepackage{verbatim}
\usepackage{mathtools}

\begin{document}

%\title{Effective potential approach to tidal locking}
\title{Tidal locking and the gravitational fold catastrophe}

\author{
Andrea Ferroglia$^{1,2}$ and Miguel C. N. Fiolhais$^{3,4}$
\\[3mm]
{\footnotesize {\it 
$^1$The Graduate School and University Center, The City University of New York, 365 Fifth Avenue, New York, NY 10016, USA \\
$^2$ Physics Department, New York City College of Technology, The City University of New York, 300 Jay Street, Brooklyn, NY 11201, USA \\
$^3$ Science Department, Borough of Manhattan Community College, The City University of New York, \\ 
     199 Chambers St, New York, NY 10007, USA \\
$^4$ LIP, Departamento de F\'{\i}sica, Universidade de Coimbra, 3004-516 Coimbra, Portugal\\
}}
}

%\date{\today}
%~\\[-1em]
\begin{abstract}
The purpose of this work is to study the phenomenon of tidal locking {in a pedagogical framework} by analyzing the effective gravitational potential of a two-body system with two spinning objects. 
%The total angular momentum of the system is assumed to be conserved, while the total mechanical energy of the system can be dissipated due to tidal friction. 
{It is shown that the effective potential of such a system is an example of a fold catastrophe. In fact, the existence of a local minimum and saddle point, corresponding to tidally-locked circular orbits, is regulated by a single dimensionless control parameter which depends on the properties of the two bodies and on the total angular momentum of the system.}
%It is shown that the effective potential may or may not develop a local minimum and a saddle point, and that those extrema correspond to  tidally locked circular orbits. This ``catastrophic'' behavior is regulated by a dimensionless control parameter which depends on the properties of the two bodies and on the total angular momentum of the system. 
{The method described in this work results in compact expressions for the radius of the circular orbit and the tidally-locked spin/orbital frequency.} The limiting case in which one of the two orbiting objects is point-like is studied in detail. 
{An analysis of the effective potential, which in this limit depends on only two parameters, allows one to clearly visualize the properties of the system.} The notorious case of the Mars' moon Phobos is presented as an example of a satellite that is past the no return point and, therefore, will not reach a stable {or unstable} tidally-locked orbit.
\end{abstract}

\maketitle

\section{Introduction}

{The understanding of tides as a consequence of the Moon-Earth interaction dates back to ancient Greece, when Seleucus of Seleucia postulated in 150 BC that the rise and fall of sea levels were directly related to the Moon.~\cite{murdin} Several centuries later, the connection between the existence of tides and the gravitational force of the Moon was proposed by Johannes Kepler in 1609.~\cite{kepler} Approximately eighty years after Kepler's insight, the mathematical study of tides as the result of gravitational forces between astronomical objects was presented in Newton's Principia~\cite{newton}, where Sir Isaac Newton used his own theory of universal gravitation to explain the mechanics of ocean tides by virtue of the gravitational attraction of the Sun and the Moon.} The joint gravitational force of the Moon and the Sun produces bulges in Earth's oceans. The Earth rotates and the land masses move toward or away from these bulges. As a consequence, the level of the oceans rises and lowers periodically. This phenomenon is very well studied and is a standard topic in undergraduate analytical mechanics courses.~\cite{taylor} In particular, the analysis of tidal forces has been the focus of several academic and educational publications over the years.~\cite{arons,gron,white,withers,koenders,butikov,razmi,massi,urbassek,pujol,ng,cregg,norsen} The presence of these frictional tides leads to the interesting effect known as tidal locking. Indeed, it is a well known fact that two astronomical bodies orbiting each other, such as the Earth and the Moon, have tidally-locked configurations. The latter are configurations in which the period of the orbital motion of the Moon around the Earth and the period of the rotation of the Moon around its axis (supposed for simplicity to be perpendicular to the plane of the Moon's orbit) are equal. Consequently, the Moon always shows the same face towards the Earth. However, this was not always the case. Upon the formation of the Moon, likely resulting from the collision between the Earth and a large astronomical body, usually referred to as Theia, the orbit of the Moon was much closer to the Earth than nowadays. Strong geological evidence supports this theory and also indicates that both the Earth and the Moon were spinning faster than today hundreds of millions of years ago.~\cite{sonett} As the tidal forces between the two objects dissipate mechanical energy of the system in the form of heat, the Moon's orbit retreats from the Earth and the spinning motion of the two objects slows down. As such, the system evolves to a state where the energy of the system reaches a minimum. Currently, the Moon is receding at a rate of $4$~cm per year from the Earth,~\cite{dickey,garrick,qin} and will continue to recede until the rotation of the Earth and the Moon around their axes  matches the orbital rotation of these objects around the common center of mass of system, \emph{i.e.} tidal locking {for both orbiting bodies}. {This was first ``predicted'' by Immanuel Kant in 1754, when he proposed that the gravitational force of the Moon would slow down Earth's rotation until the two objects would be tidally locked.~\cite{kant1} While Kant's argument was merely verbal, it did provide a correct rationale.} 

Tidal friction processes are common phenomena in the Universe as several satellites, moons, planets and stars are expected to be tidally locked to their parent objects,~\cite{makarov,zahn75,zahn77,zahn,auclair} often playing an important role on these bodies. For example, the heat resulting from the tidal friction between Jupiter and Io is believed to generate most of the geological activity on this moon.~\cite{celnikier,tyler} {Tidal forces are also relevant in the interaction between galaxies. For example, the gradient of the gravitational force is known to distort the shape of two galaxies orbiting each other.~\cite{toomre} It also possible that the Oort cloud of comets and icy planets around the Solar System is deformed by Milky Way's galactic tide.~\cite{fouchard}}  Tidal locking in two-body systems was studied in several {technical} works~\cite{counselman73,kopal72,vanhamme79,hut80} under {a variety of} different assumptions on the {initial} shape of the orbit and on the {initial} relative direction of the orbital and spin angular momenta of the two bodies.
The purpose of this work is to study tidal locking {in a didactic manner} by minimizing the effective potential of the two bodies in the problem. Even though, due to the presence of a dissipation mechanism, the mechanical energy of the system is not a constant of the problem, the total angular momentum of the system can be assumed to be constant, as  there are no external torques applied to the two-body system.~\cite{mcdonald} In this paper, the general case of two spherical rigid objects spinning  around axes perpendicular to the plane of the orbit is considered. The relevant input quantities in the problem are the masses of the two objects, the moments of inertia of the spinning objects, and  the total angular momentum of the system. By using solely the minimization of the effective potential it is possible to prove that the system can evolve toward a circular orbit in which the two objects are tidally locked. The existence of {stable} tidally-locked configurations is related to the value of a dimensionless control parameter that depends on the orbiting objects' masses and moments of inertia as well as on the total angular momentum of the system. If this control parameter is below a certain critical value, a stable, tidally-locked circular orbit configuration of the two-body system exists. {The condition on the control parameter is equivalent to the constraints on the relation between  the total angular momentum of the system and the masses and moments of inertia of the two bodies found in Counselman,~\cite{counselman73} Kopal,~\cite{kopal72} and  Hut.~\cite{hut80}}
Since the ultimate fate of the system depends on the value of a control parameter, one can say that the system exhibits a ``catastrophic'' behavior.~\cite{guemez,fiolhais} {If circular, tidally-locked configurations exist,} the final rotational (and orbital) angular velocity and the radius of the orbit are functions of the masses, moments of inertia, {total angular momentum} and of the Newton's gravitational constant $G$ only.
The minimization of the effective potential also leads to explicit compact expressions for the orbital radius and the orbital/spin angular velocity at stable and unstable tidally-locked configurations of the system.
The particular case in which one of the objects (referred to as the satellite) is considered to be point-like is studied in detail. Since in the latter limit the {shape of the} effective potential depends only on two independent dimensionless parameters, a study of {this potential} allows one to clearly visualize the location of the stable and unstable circular tidally-locked configurations.
This particular case is used to analyze the tidal interaction between Mars and its moon Phobos and their inevitable future collision.~\cite{bills,efroimsky} While this approach does not give information about the time evolution of the system, it provides a straightforward way to determine if  stable tidally-locked configurations exist. The time evolution of the system was considered for example by Hut.~\cite{hut81} 

The paper is organized as follows. The general physical situation involving two spinning objects of different masses and moments of inertia interacting through gravity is considered in Section~\ref{sec:general}. In order to simplify the analysis, the effective potential is written in terms of dimensionless variables and parameters. In Section~\ref{sec:pointlike}, the simpler situation in which one of the two objects (\emph{i.e.} 
the satellite) is taken as point-like (\emph{i.e.} it has a mass but zero moment of inertia) is studied. In particular, in Section~\ref{sec:circorb} the approximation of circular orbits is employed, while in Section~\ref{sec:elliptic} the more general case  of a point-like satellite with an elliptic orbit is considered.
Section~\ref{sec:phobos} contains an application of the approach discussed in Section~\ref{sec:circorb} to the realistic case of Mars and its moon Phobos. Conclusions are drawn in Section~\ref{sec:conclusions}.

%In the latter situation the effective potential depends only on one dimensionless parameter and on two variables; consequently, it can be easily visualized.

%is not relevant in order to find the final orbit and rotational angular velocity of the objects involved. Conversely, the angular momentum of the system is constant, as it is assumed that there are no external torques applied to it.

%assuming that the central object can spin around an axis that is perpendicular to the plane of the orbit. 
%One particular limiting cases are also studied in detail assuming circular orbits:
%\begin{itemize}
%    \item[i)] The case in which one of the objects (\emph{i.e.} the satellite) is considered to be point-like while the central object  is taken as a sphere that can spin around an axis perpendicular to the plane of the orbit. 
%    \item[ii)] The case in which both spherical objects have the same mass and moment of inertia.
%\end{itemize}

%\section{Effective potential with tidal interaction}

\section{Two body problem with two spinning objects} \label{sec:general}

\begin{figure}[t]
\begin{center}
\vspace{-1cm}
\epsfig{file=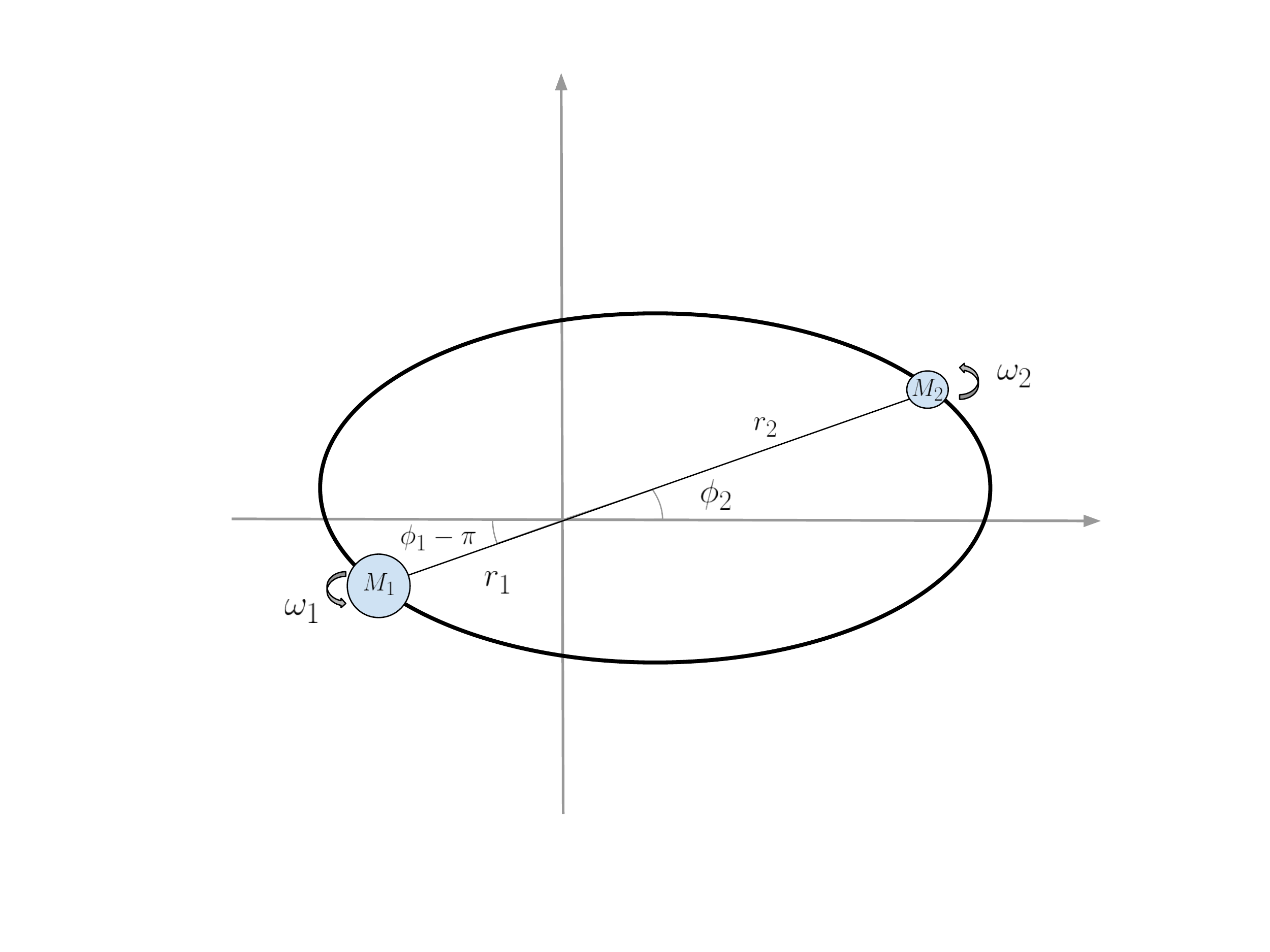,height=8.0cm}
\vspace*{-1cm}
\caption{Two spinning objects orbiting each other in system center of mass frame.}
\label{fig:config}
\end{center}
\end{figure}

This section discusses the case in which two spherical objects, of masses $M_1$ and $M_2$ and moments of inertia $I_1$ and $I_2$, respectively, are orbiting each other. The radial coordinates of the two objects in the center-of-mass frame of reference are indicated by $r_1$ and $r_2$, respectively, so that 
\begin{equation}
    M_1 r_1 - M_2 r_2 = 0 \, . \label{eq:com}
\end{equation}
The orbital angular velocities of the two bodies are indicated by $\dot{\phi}_1$ and $\dot{\phi}_2$, while the angular velocities with which they spin around their axes are indicated by $\omega_1$ and $\omega_2$, respectively.
{The configuration of the two objects in the center of mass frame is shown in Figure~\ref{fig:config}.}

The total angular momentum of the system is the sum of the orbital and spin angular momenta of both bodies, therefore
\begin{equation}
    L = M_1 r_1^2 \dot{\phi}_1  + I_1 \omega_1 + M_2 r_2^2 \dot{\phi}_2  + I_2 \omega_2 \, .
    \label{eq:angmom}
\end{equation}
Since no external torque is acting on the system, the total angular momentum is a constant of motion.
In addition, the total linear momentum of the system in the center of mass frame is zero. The conservation of {linear} momentum leads to the following conditions involving the radial and orbital velocities
\begin{align}
    M_1 \dot{r}_1 -  M_2 \dot{r}_2 &= 0 \, , \\
    M_1 \dot{\phi}_1  r_1 - M_2 \dot{\phi}_2  r_2 &= 0 \, .
\end{align}
The latter equation, in combination with angular momentum conservation and with Eq.~(\ref{eq:com}), leads to the relations \begin{equation}
    \dot{\phi}_1 = \dot{\phi}_2 = \frac{M_1 (L -I_1 \omega_1 -I_2 \omega_2 )}{M_2 (M_1+M_2) r_2^2} \, .
    \label{eq:orbitalvelocity}
\end{equation}
 In a two-body problem where the bodies are interacting through a central force, it is customary to reformulate the {total mechanical energy} of the problem in terms of a kinetic term for the radial velocity and  an effective potential depending on the distance between the two bodies.  For the two-body problem under consideration, the effective potential is the sum of the gravitational potential energy, the rotational kinetic energy for the two bodies and the usual centrifugal term proportional to the square of the orbital angular momentum and to the inverse of the square of the distance between the two objects. By systematically expressing $r_1$ in terms $r_2$ by means of Eq.~(\ref{eq:com}), the effective potential can be written as 
\begin{equation}
    U_{\textrm{eff}} = -  \frac{G M_1^2 M_2}{ (M_1 +M_2) r_2 } +
    \frac{1}{2} I_1 \omega_1^2 +\frac{1}{2} I_2 \omega_2^2 +
    \frac{M_1 (L  -I_1  \omega_1 -I_2  \omega_2)^2 }{ {2} M_2 (M_1 +M_2) r_2^2} \, .
\end{equation}
In this case the effective potential depends on three variables, namely $\omega_1$, $\omega_2$ and $r_2$. The goal is to find the local minimum of the effective potential, {if it exists}, and to verify that it corresponds to a situation in which both bodies are tidally locked, namely
\begin{equation}
    \dot{\phi}_1 = \dot{\phi}_2 = \omega_1 = \omega_2 \, . 
\end{equation}
    
It is convenient to rewrite the effective potential in terms of dimensionless variables defined as follows:
\begin{equation}
\tilde{r}_2 \equiv \frac{G M_1 M_2^2 r_2}{L^2} \, , \quad
\tilde{\omega}_1 \equiv \frac{L^3 (M_1 +M_2) \omega_1}{G^2 M_1^3 M_2^3} \, , \quad
\tilde{\omega}_2 \equiv \frac{L^3 (M_1 +M_2) \omega_2}{G^2 M_1^3 M_2^3} \, .
\end{equation}
In addition, it is useful to introduce the dimensionless parameters
\begin{equation}
    k_1 \equiv \frac{G^2 M_1^3 M_2^3 I_1}{L^4 (M_1 +M_2)} \, , \quad k_2 \equiv \frac{G^2 M_1^3 M_2^3 I_2}{L^4 (M_1 +M_2)} \, ,
\end{equation}
which are always positive. With these definitions, the effective potential becomes
\begin{equation}
    U_{\textrm{eff}} = \frac{G^2 M_1^3 M_2^3}{L^2 (M_1+M_2)} \left[\frac{1}{2 \tilde{r}_2^2} - \frac{1}{\tilde{r}_2} - \frac{k_1 \tilde{\omega}_1}{\tilde{r}_2^2} - \frac{k_2 \tilde{\omega}_2}{\tilde{r}_2^2} + \frac{k_1 \tilde{\omega}^2_1}{2} +\frac{k_2 \tilde{\omega}^2_2}{2} +  \frac{1}{2} \left(\frac{k_1  \tilde{\omega}_1 + k_2 \tilde{\omega}_2}{\tilde{r}_2} \right)^2\right] \, . \label{eq:Ueffnorm}
\end{equation}
{The rest of this section is devoted to the study of the local extrema of the effective potential in Eq.~(\ref{eq:Ueffnorm}). However, the study of this potential simplifies considerably in the case in which one of the two objects is point-like, which is considered in Section~\ref{sec:pointlike}. The reader that wants to become familiar with the simpler case first, can move directly to Section~\ref{sec:pointlike} from this point, and return to the rest of this section later on.}
By studying the solutions of the systems of equations
\begin{equation}
    \frac{\partial U_{\textrm{eff}}}{\partial \tilde{\omega}_1} =  
    \frac{\partial U_{\textrm{eff}}}{\partial \tilde{\omega}_2} =  
    \frac{\partial U_{\textrm{eff}}}{\partial \tilde{r}_2} = 0 \, , \label{eq:extremader}
\end{equation}
one can find that the local extrema of the potential must necessarily be a solution of the equation
\begin{equation}
    (\tilde{r}_2^2 + k_1 +k_2)^2 - \tilde{r}_2^3 = 0 \, . 
    \label{eq:mineq2}
\end{equation}
{A detailed derivation of Eq.~(\ref{eq:mineq2}) can be found in Appendix~\ref{app:appA}.}
Eq.~(\ref{eq:mineq2}) has four roots; two of these roots are always complex. The other two  roots  are real if
\begin{equation}
    k_1 + k_2  \le \frac{27}{256} \, , \label{eq:condks}
\end{equation}
otherwise all of the roots are complex and the effective potential does not present any local maxima, minima, nor saddle points.
If the condition in Eq.~(\ref{eq:condks}) is met, {a straightforward analysis of the Hessian determinant  shows that} one of the real roots of Eq.~(\ref{eq:mineq2}) corresponds to a minimum of the potential, while the other corresponds to a saddle point. Therefore, $k_1+k_2$ is the control parameter that determines the behavior of the system. {The condition in Eq.~(\ref{eq:condks}) is equivalent to the condition  found by Hut,~\cite{hut80} which states that the two-body system can have a stable tidally-locked configuration if the total angular momentum   satisfies the inequality
\begin{equation}
    L \ge 4 \left[\frac{G^2}{27} \frac{M_1^3 M_2^3}{M_1 + M_2} \left(I_1+I_2 \right) \right]^{\frac{1}{4}} \, .
\end{equation}
Equivalent conditions on the angular momentum derived under different sets of initial assumptions can also be found in works by Counselman,~\cite{counselman73} Kopal,~\cite{kopal72} and van Hamme.~\cite{vanhamme79}
}

{In tidally-locked configurations, where $\omega_1 = \omega_2$, the total angular momentum in Eq.~(\ref{eq:angmom}) can be rewritten as
\begin{equation}
    L = \omega_1 \left[I_1 + I_2 + \frac{M_2}{M_1} \left(M_1 + M_2 \right)r_2^2 \right]  \, , \label{eq:LLocked}
\end{equation}
{where the relation $r_1 = r_2 M_2 /M_1$, valid in the center of mass frame, was employed.}  One can rewrite Eq.~(\ref{eq:LLocked}) in terms of dimensionless quantities and  obtain that 
\begin{equation}
    \tilde{\omega}_1 = \tilde{\omega}_2 = \tilde{\omega}_\star \equiv \frac{1}{k_1+k_2 +\tilde{r}^2_\star} \, . \label{eq:localmin}
\end{equation}
where $\tilde{r}_\star$ indicates the value of $\tilde{r}_2$ corresponding to  the minimum of the effective potential.}

It must be observed that while the effective potential in Eq.~(\ref{eq:Ueffnorm}) is a function of the  two independent parameters $k_1$ and $k_2$, the coordinates of the local minimum $\tilde{r}_\star$ and $\tilde{\omega}_\star$  depend only on  a specific combination of these parameters, namely $k_1 + k_2$.  {By inserting Eq.~(\ref{eq:localmin}) in Eq.~(\ref{eq:orbitalvelocity}), it is straightforward to check that
\begin{equation}
    \dot{\phi}_1 = \dot{\phi_2} = \omega_1 = \omega_2 = \frac{G^2 M_1^3 M_2^3}{L^3 \left(M_1 + M_2 \right)} \tilde{\omega}_\star \ ,
\end{equation}
}corresponding to a tidally-locked configuration, with both objects in a circular orbit around the common center-of-mass. It is possible to find an analytical expression for $\tilde{\omega}_\star$ and $\tilde{r}_\star$ with a Computer Algebra System such as \texttt{Mathematica}.~\cite{mathematica} {These analytical expressions are long, but they can be written in a compact form by means of an appropriate reparameterization. For example, one can introduce the quantities $\epsilon$ and $u(\epsilon)$ through the relations
\begin{equation}
    k_1 + k_2 = \frac{27}{256} (1 - \epsilon^2) \, , \qquad u(\epsilon) = (1-\epsilon)^{\frac{2}{3}} (1 + \epsilon)^{\frac{1}{3}} + 
    (1-\epsilon)^{\frac{1}{3}} (1 + \epsilon)^{\frac{2}{3}} \, .
    \label{eq:k1+k2param}
\end{equation}
When written in terms of these quantities, the normalized frequency and radius corresponding to the local minimum, {(which are solutions of the system in Eq.~(\ref{eq:extremader}) and, consequently, satisfy Eq.~(\ref{eq:mineq2}))} become
\begin{align}
    \tilde{\omega}_\star =& \frac{32}{27 (1- \epsilon^2)} \left( 6 - \sqrt{4 +6 u} - \sqrt{8 - 6 u + \frac{8 \sqrt{2}}{\sqrt{2 +3 u}}}  \right) \, ,\label{eq:oms1}  \\ 
    \tilde{r}^2_\star =& -\frac{27}{256} (1 - \epsilon^2) \left[1 + 8 \left(- 6 + \sqrt{4 +6 u} + \sqrt{8 -6 u +\frac{8 \sqrt{2}}{\sqrt{2 +3 u}} }  \right)^{-1} \right] \, . \label{eq:rstar}
\end{align}
Also the coordinates  of the saddle point, indicated with $\tilde{\omega}_s$ and $\tilde{r}_s$, {(which are also solutions of the system in Eq.~(\ref{eq:extremader}))} have similar compact expressions
\begin{align}
    \tilde{\omega}_s =& \frac{32}{27 (1- \epsilon^2)} \left( 6 - \sqrt{4 +6 u} + \sqrt{8 - 6 u + \frac{8 \sqrt{2}}{\sqrt{2 +3 u}}}  \right) \, , \label{eq:oms2}  \\
    \tilde{r}^2_s =& -\frac{27}{256} (1 - \epsilon^2) \left[1 + 8 \left(- 6 + \sqrt{4 +6 u} - \sqrt{8 -6 u +\frac{8 \sqrt{2}}{\sqrt{2 +3 u}} }  \right)^{-1} \right] \, . \label{eq:rs}
\end{align}
{It should be observed that by choosing to employ $\epsilon^2$ rather than $\epsilon$ in the parameterization of $k_1 +k_2$ in Eq.~(\ref{eq:k1+k2param}), it becomes possible to factor $1- \epsilon^2 = (1-\epsilon) (1+\epsilon)$. This helps in obtaining compact expressions since the two factors appear at different powers in the auxiliary function $u(\epsilon)$. }
}
\begin{figure}[t]
\begin{center}
\epsfig{file=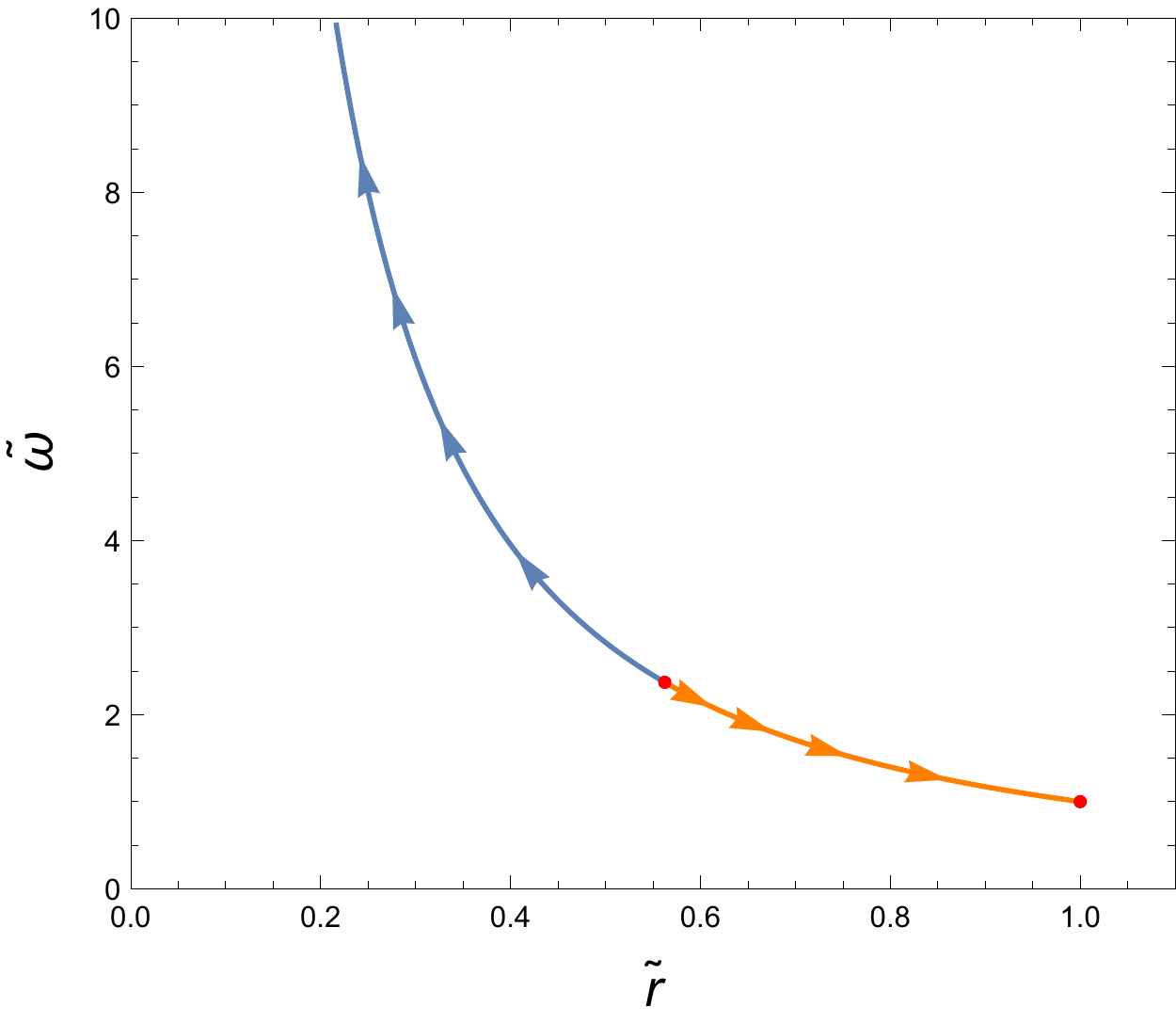,height=9.0cm}
\caption{{Location of the saddle point (in blue) and of the local minimum (in orange) in the $\{\tilde{r},\tilde{\omega}\}$ plane {as the value of the control parameter varies in the range $0 \le k_1+k_2 \le 27/256$}. The arrows indicate the direction of the change in the location of the points as the parameter $\epsilon$ is varied between $0$ (\emph{i.e.} the critical value) and $1$. The common location of the local minimum and of the saddle point for $\epsilon =0$ is indicated by the red dot between the blue and orange lines; its coordinates are $\{9/16,64/27\}$. At $\epsilon = 1$ the local minimum is found at the end of the orange line, located at $\{1,1\}$.}}
\label{fig:sadmin}
\end{center}
\end{figure}

{
By using Eqs.~(\ref{eq:oms1},\ref{eq:rstar},\ref{eq:oms2},\ref{eq:rs}), it is possible to plot the location of the saddle point and the local minimum in the
$\{\tilde{r},\tilde{\omega}\}$  plane {as the value of the control parameter varies}. The set of points where the saddle point and local minimum can be located is shown in Figure~\ref{fig:sadmin}, where the blue line indicates the locus of the saddle point and the orange line indicates the locus the local minimum. The arrows indicate the direction of the position change of the points as the parameter $\epsilon$ is varied from $\epsilon = 0$, corresponding to the critical value $k_1+k_2 = 27/256$, to $\epsilon = 1$, which corresponds to $k_1+k_2 = 0$. The saddle point reaches arbitrarily large values of $\tilde{\omega}$ as $\epsilon \to 1$, while in the same limit the local minimum approaches the point of coordinates $\{1,1\}$. 
}

\section{Two body problem with point-like satellite \label{sec:pointlike}}

At this stage, it is interesting to analyze the special case in which one of the two objects described in the previous section has a mass that is negligible in comparison to the mass of the other object. The two objects will be referred to as the satellite and the central object, respectively. The mass of the central object will be relabeled as $M_1=M$ and its moment of inertia as $I_1=I$. Similarly, the point-like satellite mass is now $M_2=m$ and its moment of inertia is neglected, \emph{i.e.} $I_2=0$. {This is equivalent to neglecting the tidal interaction between the central object and the structure of the satellite, and therefore the angular speed of satellite $\omega_2$ is no longer part of this analysis. {In the case considered in this section, tidal locking occurs when the satellite's orbital angular speed around the central object is equal to the angular speed of the central object around its axis.}} As the mass of the central object is assumed to be much larger than the satellite's mass, {$m \ll M$} , the position of the satellite is written as $r_2=r$, while $r_1=0$, and the dimensionless parameter associated with the central object's moment of inertia is now expressed as
\begin{equation}
    k \equiv \frac{G^2 M^3 m^3 I}{L^4 (M + m)}  {\,\,\, \xrightarrow[]{m \ll M} \,\,\,}\frac{G^2 M^2 m^3 I}{L^4} \, .
\end{equation}
In this notation, the effective potential of the system becomes,
\begin{equation}
    U_{\textrm{eff}}(\tilde{r},\tilde{\omega}) \equiv \frac{G^2 M^2 m^3}{L^2} \tilde{U}_{\textrm{eff}}(\tilde{r},\tilde{\omega}) = \frac{G^2 M^2 m^3}{L^2} \left[\frac{1}{2 \tilde{r}^2} - \frac{1}{\tilde{r}} - \frac{k \tilde{\omega}}{\tilde{r}^2} + \frac{k \tilde{\omega}^2}{2} +  \frac{1}{2} \frac{k^2  \tilde{\omega}^2}{\tilde{r}^2} \right] \, , \label{eq:Ueffsat}
\end{equation}
with
\begin{equation}
\tilde{r} \equiv \frac{G M m^2 r}{L^2} \, , \quad
\tilde{\omega} \equiv \frac{L^3 \omega}{G^2 M^2 m^3} \, ,
\end{equation}
{where $\omega$ is the rotational frequency of the planet around its axis} and $\tilde{U}_{\textrm{eff}}$ is the dimensionless effective potential.
As it can be seen from the discussion in the previous section, the potential in Eq.~(\ref{eq:Ueffsat}) presents a local minimum and a saddle point if $k < 27/256$. 
 
\subsection{{Circular orbit}} \label{sec:circorb}
 
 It is interesting to see what one can learn by imposing that the orbit of the point-like satellite around the massive planet remains always circular. {This particular configuration was already considered more than 150 years ago by Darwin.~\cite{darwin79} Since under these assumptions the effective potential depends on a single variable, the results derived in the previous section can be easily visualized and interpreted.}
The condition that ensures that the orbit is circular is 
\begin{equation}
\dot{\phi}^2 = \frac{G M}{r^3} \, , 
\end{equation}
which coincides with Kepler's third law in the case of a circular orbit. Consequently the {conserved} angular momentum can be written as
\begin{equation}
    L = m r^2 \sqrt{ \frac{G M}{r^3} } +     I \omega \, . \label{eq:circ}
\end{equation}
If rewritten in terms of normalized quantities, the relation in Eq.~(\ref{eq:circ}) leads to the condition %
\begin{equation}
    \tilde{\omega} = \frac{1 - \sqrt{\tilde{r}}}{k} \, . \label{eq:circnorm}
\end{equation}
By imposing the relation in Eq.~(\ref{eq:circnorm}) between $\tilde{\omega}$ and $\tilde{r}$ in the effective potential one finds
\begin{equation}
    U_{\textrm{eff}}(\tilde{r}) \equiv \frac{G^2 M^2 m^3}{L^2} \tilde{U}_{\textrm{eff}}(\tilde{r}) = \frac{G^2 M^2 m^3}{L^2} \left[ \frac{1}{2 k} - \frac{1}{2 \tilde{r}} - \frac{\sqrt{\tilde{r}}}{k} + \frac{\tilde{r}}{2 k}\right] \, . \label{eq:Ueffcirc}
\end{equation}

\begin{figure}[t]
\begin{center}
\epsfig{file=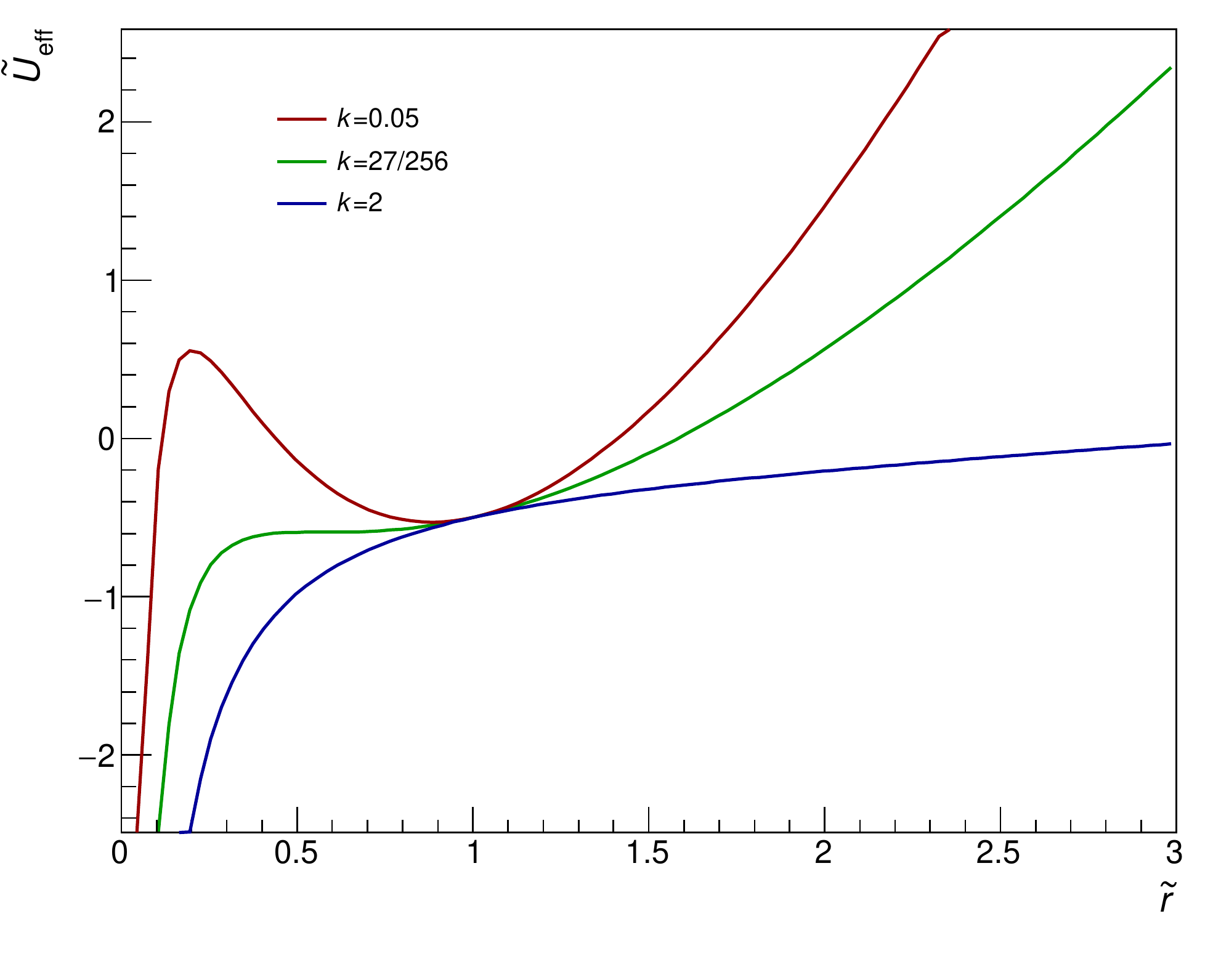,height=10.0cm}
\caption{Dimensionless effective potential defined in Eq.~(\ref{eq:Ueffcirc}) for $k$ below, above, and equal to the critical value $27/256$. {The critical value of $k$ corresponds to the green line, where the maximum and the minimum meet at the same value of $\tilde{r}$ and annihilate.}}
\label{fig:normalizedU}
\end{center}
\end{figure}

Figure~\ref{fig:normalizedU} shows the dimensionless potential, $\tilde{U}_{\textrm{eff}}(\tilde{r})$, in Eq.~(\ref{eq:Ueffcirc}) for different values of the parameter $k$. For $k < 27/256$ this potential has a local maximum and a local minimum, which can be observed in the red line of the figure. The maximum corresponds to the saddle point in the potential in Eq.~(\ref{eq:Ueffsat}), where the condition of a circular orbit was not imposed. The minimum of the potential in Eq.~(\ref{eq:Ueffcirc}) corresponds to the minimum of the potential in Eq.~(\ref{eq:Ueffsat}). For $k > 27/256$ the local maximum and minimum disappear, \emph{i.e.} the potential does not present any equilibrium configurations. This can be observed in the blue line in the figure. The critical {line that separates curves with a local minimum from curves without a local minimum} occurs at $k = 27/256$, represented by the green line in the figure. {In the green curve} the local maximum and minimum merge into an inflection point {with a horizontal tangent}. {This phenomenon can be interpreted as a fold catastrophe. In catastrophe theory, a fold catastrophe represents a system where a single control parameter can cause equilibria to appear or disappear.~\cite{zeeman,gilmore,saunders,poston} This behavior can be better visualized in a potential of the form $U(x) = x^3 + b x$, where $b$ represents a real number. For $b<0$, the potential displays stable and unstable equilibrium solutions. However, at the critical value $b=0$, the maximum and the minimum meet and annihilate each other. For $b>0$, there are no equilibrium solutions. In the case presented in Figure~\ref{fig:normalizedU}, the dimensionless $k$ acts as a control parameter of the fold catastrophe.}  One can also observe that for $\tilde{r}=1$, the dimensionless potential in Eq.~(\ref{eq:Ueffcirc})
becomes independent from $k$ and equal to $1/2$. For this reason the three curves in Figure~\ref{fig:normalizedU} coincide at $\tilde{r}=1$.

Both the maximum and the minimum of the potential in Eq.~(\ref{eq:Ueffcirc}) correspond to tidally-locked configurations. Indeed, one is in the presence of a tidally-locked configuration whenever
\begin{equation}
    \dot{\phi} = \omega \, .
\end{equation}
This implies that in a tidally-locked configuration the fixed angular momentum can be expressed as
\begin{equation}
    L = \omega \left(I + m r^2 \right) \, . \label{eq:Llocked}
\end{equation}
In turn, the relation in Eq.~(\ref{eq:Llocked}) can be written in terms of the normalized quantities as
\begin{equation}
   \tilde{\omega} = \frac{1}{  k + \tilde{r}^2 } \,. \label{eq:tidallock}
\end{equation}
{As expected, Eq.~(\ref{eq:tidallock}) is equivalent to Eq.~(\ref{eq:localmin}) in the limit in which $k_2 \to 0$, since in the case of a point-like satellite there is a single moment of inertia and consequently, a single parameter $k$.}
With the help of a Computer Algebra System, it is straightforward to verify that at the minimum and at the maximum of the potential in Eq.~(\ref{eq:Ueffcirc}), the relations in Eq.~(\ref{eq:circnorm}) and in Eq.~(\ref{eq:tidallock}) are simultaneously verified, namely
\begin{equation}
  \frac{1 - \sqrt{\tilde{r}}}{k} = \frac{1}{  k + \tilde{r}^2 } \, . \label{eq:simcond}
\end{equation}
Also, the analytical expressions for the solutions of the quartic  Eq.~(\ref{eq:simcond}) for  $\tilde{r}$ can be easily found by using {\tt Mathematica}. {As in the general case discussed in Section~\ref{sec:general},} 
Eq.~(\ref{eq:simcond}) has two real solutions for $k < 27/256$. {The analytical expressions in Eqs.~(\ref{eq:oms1},\ref{eq:rstar},\ref{eq:oms2},\ref{eq:rs}) are valid also in the special case discussed in this section, {with $k = 27/256 (1-\epsilon^2)$.}}

It must be stressed that the dissipative forces between the two objects are not included in this model. As such, even though the derivative of this effective potential corresponds to an effective force, the time-evolution of the system cannot be obtained from this potential alone because the dissipative tidal forces are not taken into account. {In order to derive the tidal force, it would be necessary to perform a detailed calculation of the gradient of the gravitational force that the objects apply on the extended shape of each other.} As the system progresses from a higher energy configuration to a lower energy state, the difference in the effective potential energy is dissipated in the form of heat and radiation due to these non-conservative forces. This is analogous to the case of the damped simple harmonic motion where the quadratic potential alone only provides the possible energy configurations of the system but not its dynamics, unless the damping force is specified. It should be observed once more that, unlike energy, the total angular momentum of the system remains constant in this analysis; in fact, all the forces in the system are assumed to be internal, the total net torque is zero and angular momentum is conserved.

In Figure~\ref{fig:normalizedU}, the effective potential of the system for $k=0.05$ clearly shows two points of equilibrium, an unstable one closer to the central object and a stable one farther away. As observed above, both these states correspond to tidally-locked configurations, \emph{i.e.} geostationary orbits (if the central object is the Earth). The unstable equilibrium will be indicated with $R_{\textrm{max}}$ and the stable one with $R_{\textrm{min}}$. 

On the one hand, if the satellite is between the surface of the planet and $R_{\textrm{max}}$, the energy of the system is always smaller than the local maximum of the effective potential. This is the no return point, where the satellite will inevitably decay its orbit until it collides with the planet. This is known as tidal deceleration, where objects within geostationary orbits move faster than the spinning motion of the Earth. Consequently, the tide produced by the satellite dissipates the energy of the system progressively decreasing the radius of its orbit. The moon of Mars, Phobos, is a well known example of such a state, where the orbit of the moon is within the areosynchronous orbit (\emph{i.e.} geosynchronous for Mars). 
On the other hand, if the satellite is in an orbit with a radius larger than $R_{\textrm{max}}$, the dissipation mechanism will bring the satellite to a circular orbit with radius $R_{\textrm{min}}$, where the dissipative forces cease to do work and the systems remains in equilibrium.

%{As a possible in-class activity, students could be challenged to analyze the effective potential for the other moon of Mars, Deimos. For example, students could solve Eq.~(\ref{eq:simcond}) numerically for this system, in order to identify the real solutions for the stable and unstable orbits of Deimos. Other systems where the satellite size is not completely negligible, like the Earth and the Moon or Pluto and Charon, could be used to test the limitations of this approximation. Another possibility would be to ask the students to plot the effective potential in Eq.~(\ref{eq:Ueffnorm}) for several values of the variable $k$, similarly to Figure~\ref{fig:normalizedU}. The plot can be  turned into an animation where the value of $k$ is changing in time by using the {\tt Mathematica} function {\tt Manipulate}.~\cite{mathematica}}

\subsection{The case of Phobos \label{sec:phobos}}

\begin{table}[t]
\begin{center}
\begin{tabular}{|c c|c c| }
\hline
 Mars mass & $6.39 \times 10^{23}$~kg &
 Phobos mass & $1.08 \times 10^{16}$~kg \\  
 Mars radius & $3.39 \times 10^6$~m  & Mars moment of inertia & $2.96 \times 10^{36}$~kg m$^2$\\ 
 Radius of Phobos orbit & $9.37 \times 10^6$~m & Phobos period & $2.75 \times 10^4$~s \\
Phobos orbital angular velocity  & $2.28 \times 10^{-4}$~rad/s & Mars spin angular velocity&  $7.09 \times 10^{-5}$~rad/s\\
Mars-Phobos system $L$ & $2.08 \times 10^{32}$~kg m$^2$/s & $k$ & 
$3.57\times 10^{-18}$ 
\\
Current $\tilde{r}$ for Phobos & 
$1.07 \times 10^{-12}$ 
 &$\tilde{r}_{\text{max}}$& $2.34 \times 10^{-12}$\\
$\tilde{r}_{\text{min}}$ & 
$1$ 
 & & \\
\hline 
\end{tabular}
\caption{Parameters relevant for the description of the Mars-Phobos system. \label{tab:marsphobos}}
\end{center}
\end{table}

\begin{figure}[h]
\begin{center}
\epsfig{file=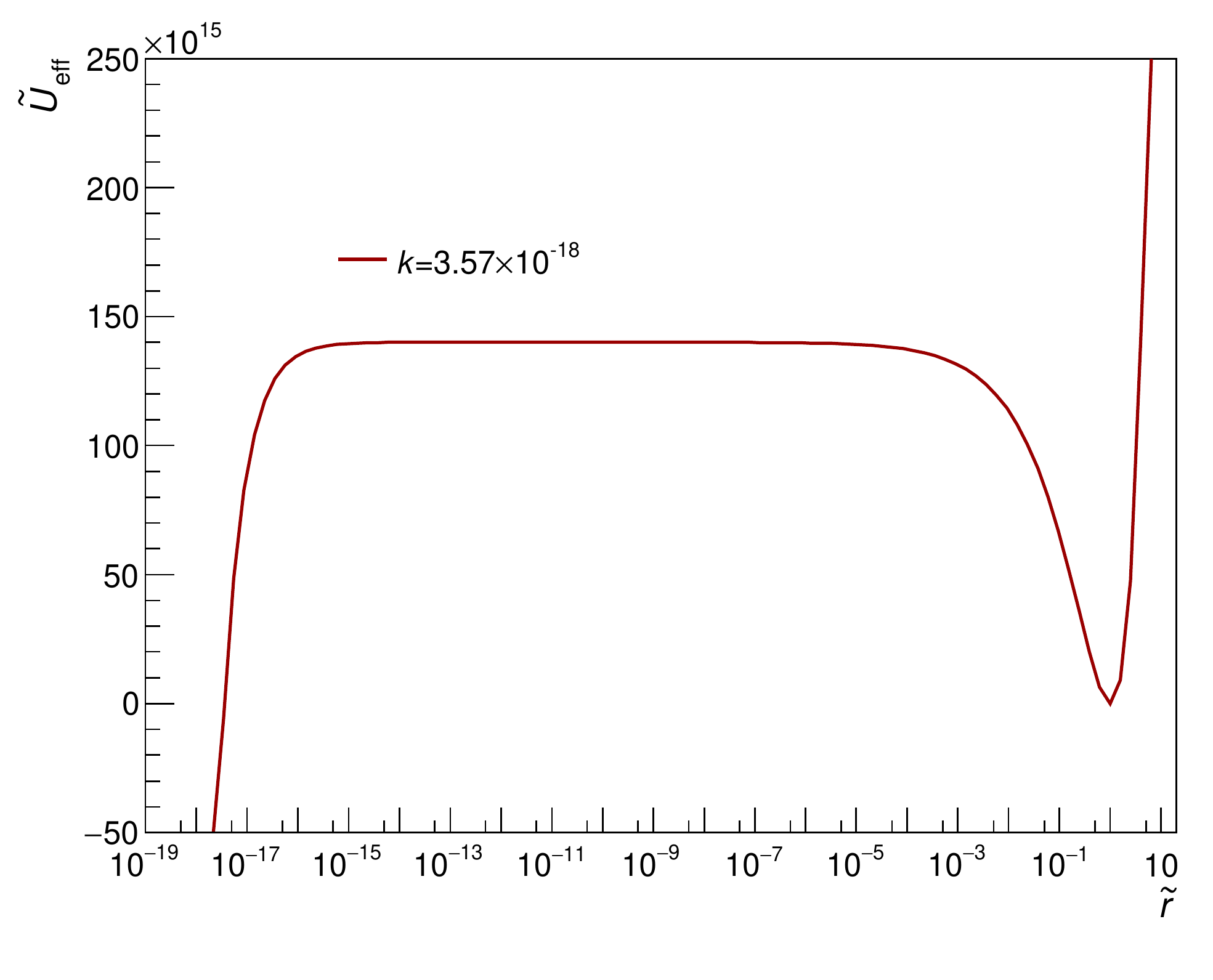,height=6.5cm}
\epsfig{file=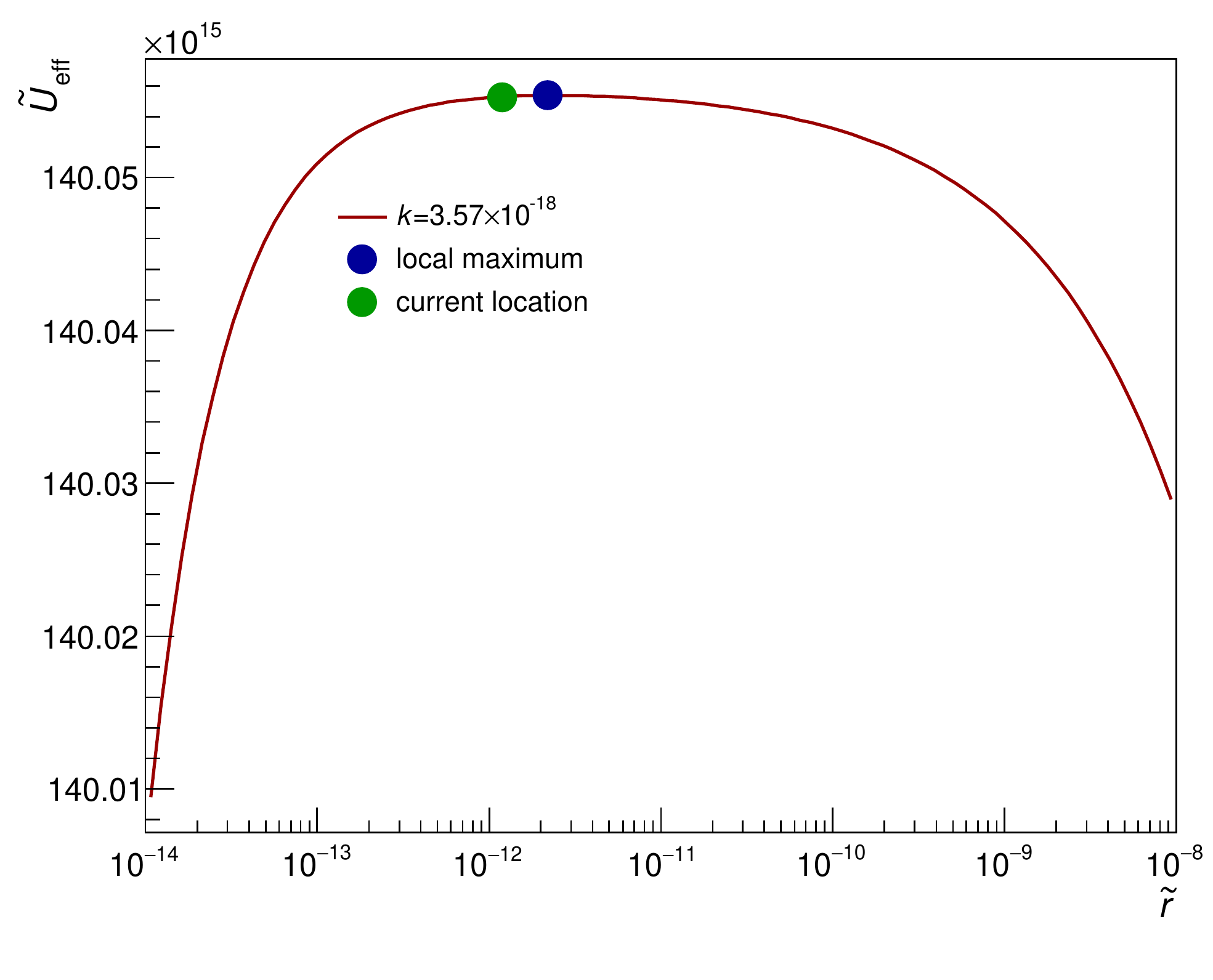,height=6.5cm}

\caption{The left panel shows the dimensionless effective potential for the Mars-Phobos system. The right panel shows in greater detail the region around the maximum.}
\label{fig:PhobosMars}
\end{center}
\end{figure}

{The orbit of Mars' moon Phobos} lends itself to be analyzed in the context of the discussion carried out in Section~\ref{sec:circorb}. This moon orbits Mars with a small eccentricity, so that the circular orbit approximation is justified.  The relevant parameters for the Mars-Phobos system are listed in Table~\ref{tab:marsphobos}. Phobos' is currently within the areostationary  orbit, because its orbital period is shorter than Mars' day. As in the previous discussions, any interaction  between the Phobos-Mars system and the Sun, other planets and moons is ignored. The dimensionless effective potential for the Phobos-Mars system is shown in Fig~\ref{fig:PhobosMars}. One can see that the current position of Phobos is to the left of the maximum, which corresponds to the areostationary orbit. Consequently, as the Phobos-Mars system loses energy through tidal friction, Phobos will get closer and closer to Mars and eventually will fall onto it. This phenomenon is referred to as tidal deceleration. If the satellite was located past the areostationary orbit (\emph{i.e.} if the satellite had an orbital period longer than  Mars' day, like the case of Deimos) then it would be pushed away from Mars toward the minimum of the dimensionless potential. As the mass of Phobos is much smaller than the mass of Mars, the parameter $k$ is much smaller than $1$. Consequently, the minimum of the dimensionless potential is located at $\tilde{r}_{\text{min}} \sim 1$. This corresponds to a physical distance from Mars of the order of $10^{19}$~m, well beyond the edges of the solar system. This process is known as tidal acceleration.

\subsection{{Elliptic orbit} \label{sec:elliptic}}

\begin{figure}[ht]
\begin{center}
\epsfig{file=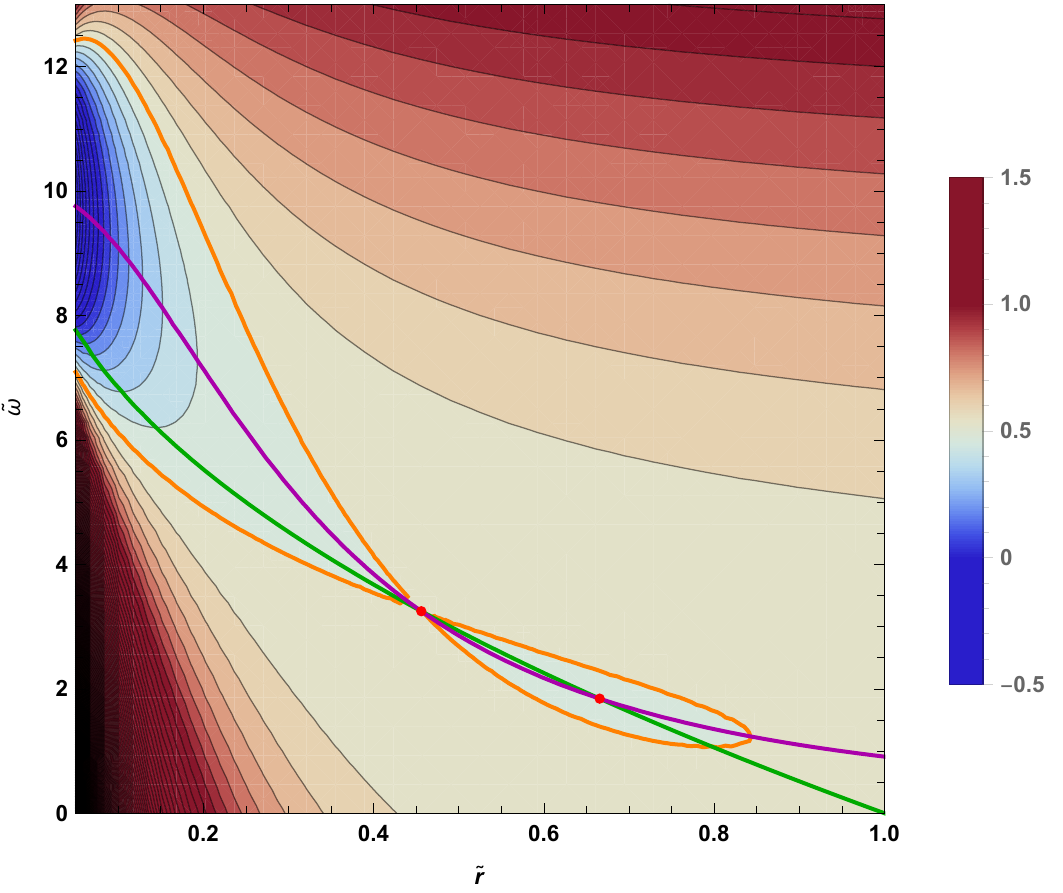,height=10.0cm}
\caption{Dimensionless effective potential defined in Eq.~(\ref{eq:Ueffsat}) for $k = 0.1$. The red points indicate the saddle point \{0.456093,3.24654\} and the minimum \{0.665617,1.84146\}. The green curve corresponds to points in the $\{\tilde{r},\tilde{\omega}\}$ plane where the orbit is tidally locked. The magenta curve corresponds to points where the satellite's orbit is circular. It must be observed that the saddle point and the minimum are the only two points in which the orbit is simultaneously tidally locked and circular. The orange line indicates points that have the same value of $\tilde{U}_{\text{eff}}$ as the saddle point. }
\label{fig:normalizedU2D}
\end{center}
\end{figure}

It is instructive to plot the dimensionless effective  potential {$\tilde{U}_{\text{eff}}$} in Eq.~(\ref{eq:Ueffsat}), {which depends on the two variables $\tilde{r}$ and $\tilde{\omega}$}, for a fixed value of $k$. 
{Since in this section $\tilde{r}$ and $\tilde{\omega}$ are considered independent variables and they not forced to satisfy Eq.~(\ref{eq:circnorm}), in general the satellite follows an elliptic orbit.} {The potential {$\tilde{U}_{\text{eff}}$} is plotted in Figure~\ref{fig:normalizedU2D}}, where the elevation of the effective potential for $k=0.1$ at a given value of $\{\tilde{r},\tilde{\omega}\}$ is shown through a contour plot and a color code. Bluish colors indicate low values of the potential while reddish colors indicate high values of the potential. 
The saddle point (located at $\{0.456093,3.24654\}$) and the local minimum (located at $\{0.665617,1.84146\}$) are indicated in the figure by red dots. The curve of all of the points where the potential has the same value as the value of the potential at the saddle point is shown in orange. As expected, this curve self-intersects at the saddle point and forms a closed loop around the local minimum.  {If the system finds itself at a value of  $\{\tilde{r},\tilde{\omega}\}$ that is within the orange loop surrounding the local minimum, further energy loss through the tidal interaction will bring the system to the local minimum. On the contrary, if the system finds itself at a point in the 
 the region bound by the two branches of the orange curve and the $\tilde{\omega}$ axis, where $\tilde{r}$ is smaller than $\tilde{r}$ at the saddle point, the system evolves toward the blue region and the satellite will eventually fall on the central object. } The set of all of the points that correspond to circular orbits also forms a continuous curve, which is indicated in magenta in the figure. The curve corresponding to the points in which there is {synchronism between rotation and revolution} is instead represented in green. It must be observed that the magenta and green curves intersect in two points, which coincide with the saddle point and the local minimum. 

There is a region of the $\{\tilde{r},\tilde{\omega}\}$ plane where the effective potential has lower values than in the local minimum. As a matter of fact the effective potential is unbound from both below and above along the line $\tilde{r} = 0$. However, that line has no direct physical meaning because $\tilde{r}$ must be larger than a certain minimum value if one requires that the {point-like} satellite is outside the central object. Indeed, the condition that ensures that the satellite is outside the central object is
\begin{equation}
    r \ge R \, ,
\end{equation}
where $R$ is the central object's radius. In terms of dimensionless quantities, this translates into the condition
\begin{equation}
    \tilde{r}^2 \ge \frac{5}{2} C k \, , \qquad \text{where} \qquad C \equiv \frac{m (M+m)}{M^2} \, .
\end{equation}
{
It is then possible to solve the system of equations
\begin{equation}
    \frac{1 - \sqrt{\tilde{r}}}{k}   = \frac{1}{  k + \tilde{r}^2 } \, , \qquad\tilde{r}^2 = \frac{5}{2} C k \, , \label{eq:sysout}
\end{equation}
with respect to $\tilde{r}$ and $C$ and for fixed $k$. Two of the system solutions coincide with the values of $\tilde{r}$ for the local minimum and the saddle point, since one of the equations in the system is Eq.~(\ref{eq:simcond}). For a given $k$ and for a value of  $\tilde{r}_0$ that satisfies Eq.~(\ref{eq:simcond}),
the second of Eqs.~(\ref{eq:sysout}) determines a  value of $C$, {indicated by $C_0$. For any value of $C$ in the range $0 \le C \le C_0$ the solution $\tilde{r}_0$ will be outside the central object}. Therefore,  since the values of $\tilde{r}$ that satisfy the first Eqs.~(\ref{eq:sysout}) are the saddle point and the local minimum, one can easily check if for a given $C$ and $k$ the saddle point or the local minimum  fall inside or outside the central object.}

\section{Conclusions}
\label{sec:conclusions}
This work studies the effective gravitational potential of two orbiting objects spinning around an axis perpendicular to the orbital plane. Assuming that no external forces are acting on the two-body system, but that mechanical energy can be dissipated through tidal forces, it is possible to show {in a pedagogical and straightforward way} that the gravitational effective potential of the system develops a local minimum if the dimensionless parameter $k_1 + k_2$ is smaller than the critical value $27/256$. 
The local minimum corresponds to a circular orbit in a tidally-locked configuration, \emph{i.e.} a configuration in which the orbital angular velocity of the objects around the system center of mass is equal to the angular velocity with which the two objects spin around their axes. For values of $k_1+k_2$ larger than the critical value, the effective potential does not have a local minimum or any sort of local extrema. 
{This result is equivalent to a known condition on the total angular momentum of the system that must be satisfied in in order to have stable configurations in which there is sychronism between revolution and rotation of the two bodies.~\cite{counselman73,kopal72, vanhamme79,hut80}} An analytical, compact expression for the radius of the circular orbit and for the orbital angular velocity at the local minimum  of the effective potential was derived in this work. {A second circular tidally-locked configuration exists, which is, however, unstable. The  radius of the circular orbit and  the orbital angular velocity at this unstable configuration are also provided. The existence of stable tidally-locked configuration is regulated by the value of one single parameter and this behavior can be interpreted as a fold catastrophe.}

The analysis becomes particularly transparent {and easy to visualize} when one considers a point-like satellite orbiting a central object that is considerably more massive than the satellite. This paper discussed in detail both the case in which the satellite's orbit is assumed to be circular and the case in which the orbit can be elliptic. In both cases, it is possible to see graphically how the presence in the effective potential of a local minimum and a local maximum (for the circular orbit case) or a  saddle point (for the case of elliptic orbits) is driven by the value of a single dimensionless parameter. This method, when applied to Mars' moon Phobos, allows one to conclude that, as already known, Phobos is bound to fall on Mars.

The method described in this work relies exclusively on the use of the effective potential for the two-body problem, {and on} basic concepts {in} rotational dynamics and mathematical analysis of multivariate functions. While the method cannot describe the dynamics of the system, it provides a simple test to see if a two body system can evolve to a stable configuration where there is synchronism between rotation and revolution and a way to obtain the angular velocity and radius of the stable configuration. The method can then have a didactical use in an introductory mechanics class.

\acknowledgements
{The authors would like to thank Pierre Auclair-Desrotour, Giovanni Ossola and Christopher Clouse for useful discussions and comments and suggestions on this manuscript.}

\appendix
\section{Polynomial equation satisfied by the local extrema \label{app:appA}}

{
In this appendix it is shown that the extrema of the effective potential in Eq.~(\ref{eq:Ueffnorm}) must satisfy the polynomial relation in Eq.~(\ref{eq:mineq2}). One of the relations that the  local extrema should satisfy is
{
\begin{equation}
    \frac{\partial U_{\textrm{eff}}}{\partial \tilde{\omega}_2} = \frac{G^2 M_1^3 M_2^3}{L^2 (M_1+M_2)} k_2 \left[ \tilde{\omega}_2 + \frac{k_1 \tilde{\omega}_1}{\tilde{r}^2_2}-\frac{1}{\tilde{r}^2_2} + \frac{k_2 \tilde{\omega}_2}{\tilde{r}^2_2} \right] = 0\, .
\label{app:interm1}
\end{equation}}
By solving Eq.~(\ref{app:interm1}) with respect to $\tilde{\omega}_2$ one finds
{
\begin{equation}
    \tilde{\omega}_2 = \frac{1 - k_1 \tilde{\omega}_1}{k_2 + \tilde{r}^2_2} \, .\label{app:interm2}
\end{equation}}}

{
Eq.~(\ref{app:interm2}) defines a surface in the three dimensional space spanned by the variables $\{\tilde{\omega}_1,\tilde{\omega}_2,\tilde{r}_2\}$.
One can then insert the condition in Eq.~(\ref{app:interm2}) in Eq.~(\ref{eq:Ueffnorm}). In this way, one finds a new expression for the effective potential valid for the points on the surface defined by   Eq.~(\ref{app:interm2}), namely,
\begin{equation}
  U^{S}_{\textrm{eff}} = \frac{1}{2} \frac{G^2 M_1^3 M_2^3}{L^2 (M_1+M_2)} \left[ \frac{k_2}{ \tilde{r}_2} \frac{k_1 \tilde{r}_2 \tilde{\omega}_1^2 -2 }{k_2 + \tilde{r}_2^2}  + \frac{k_1 \tilde{r}_2^2 \tilde{\omega}_1^2 - 2 \tilde {r}_2 + \left(k_1 \tilde{\omega}_1-1 \right)^2}{k_2 + \tilde{r}_2^2} \right] \, , 
\label{app:interm3}
\end{equation}
where the superscript $S$ was added to the l.h.s. of Eq.~(\ref{app:interm3}) to explicitly indicate that the condition in Eq.~(\ref{app:interm2}) was imposed. }

{
One of the conditions that the local extrema of Eq.~(\ref{app:interm3}) have to satisfy is 
\begin{equation}
    \frac{\partial U^{S}_{\textrm{eff}}}{\partial \tilde{\omega}_1} = \frac{G^2 M_1^3 M_2^3}{L^2 (M_1+M_2)} \frac{k_1}{k_2 + \tilde{r}_2^2} \left(\tilde{r}_2^2 \tilde{\omega}_1 +k_1 \tilde{\omega}_1 + k_2 \tilde{\omega}_1 -1\right) = 0 \, .\label{app:interm4}
\end{equation}
Eq.~(\ref{app:interm4}) is satisfied when
\begin{equation}
    \tilde{\omega}_1 = \frac{1}{k_1 + k_2 + \tilde{r}_2^2} \, .
    \label{app:interm5}
\end{equation}
In turn, when combining Eq.~(\ref{app:interm2}) and Eq.~(\ref{app:interm5}) one finds that at the local extrema of the effective potential
{
\begin{equation}
    \tilde{\omega}_1 = \frac{1}{k_1 + k_2 + \tilde{r}_2^2}  = \tilde{\omega}_2\, .
    \label{app:interm6}
\end{equation}}
Eq.~(\ref{app:interm6}) defines a curve in the three-dimensional space spanned by the variables
$\{\tilde{\omega}_1,\tilde{\omega}_2,\tilde{r}_2\}$.
Notice that Eq.~(\ref{app:interm6}), which is satisfied by the local extrema of the effective potential, is equivalent to Eq.~(\ref{eq:localmin}).
Next, by inserting the conditions in Eq.~(\ref{app:interm6}) in Eq.~(\ref{eq:Ueffnorm}), one finds an expression for the effective potential valid for the points that satisfy Eq.~(\ref{app:interm6}),
\begin{equation}
    U^{C}_{\textrm{eff}} = \frac{1}{2} \frac{G^2 M_1^3 M_2^3}{L^2 (M_1+M_2)} \frac{1}{\tilde{r}_2} \frac{\tilde{r}_2 - 2 k_1 -2 k_2 - 2 \tilde{r}_2^2 }{k_1 +k_2 + r_2^2} \, . 
\end{equation}
Also in this case, the superscript $C$ was added to indicate that the conditions in Eq.~(\ref{app:interm6}) were imposed.}

{Finally, the values of $\tilde{r}_2$ that correspond to extrema of $U_{\textrm{eff}}$ should also be the values that lead to the local extrema of $U^{C}_{\textrm{eff}}$. These values must satisfy the  condition
\begin{equation}
    \frac{\partial U^{C}_{\textrm{eff}}}{\partial \tilde{r}_2} = \frac{G^2 M_1^3 M_2^3}{L^2 (M_1+M_2)}  \frac{\left(\tilde{r}_2^2 + k_1 +k_2 \right)^2 - \tilde{r}_2^3}{\tilde{r}_2^2 \left(\tilde{r}_2^2 + k_1 +k_2 \right)^2} =0 \, , \label{app:int7}
\end{equation}
where the numerator of the second fraction coincides with Eq.~(\ref{eq:mineq2}). Consequently, the condition in Eq.~(\ref{app:int7}) is fulfilled by the solutions of Eq.~(\ref{eq:mineq2}).}

\end{document}